\newcommand{\bPsi}{\bm \Psi}

\newcommand{\tomega}{\tilde{\omega}}
\documentclass[12pt]{iopart}

\usepackage{graphicx}
\usepackage{graphics}
\usepackage{epsfig}
\usepackage{bm}
\usepackage{eurosym}
\usepackage{color}
\usepackage[normalem]{ulem}
\usepackage{amsthm}

\begin{document}

\title{Four-wave mixing in spin-orbit coupled Bose-Einstein condensates}

\author{Nguyen Viet Hung$^{1}$, Piotr Sza\'{n}kowski$^2$, Vladimir V. Konotop$^3$, and  Marek Trippenbach$^4$}

\address{$^1$Advanced Institute for Science and Technology,
Hanoi University of Science and Technology, Hanoi, Vietnam.
    \\
$^2$Institute of Physics, Polish Academy of Sciences,
al.Lotnik\'{o}w 32/46, PL 02-668 Warsaw, Poland.
    \\
$^3$Departamento de F\'isica and Centro de F\'isica Te\'orica e
Computacional, Faculdade de Ci\^encias, Universidade de Lisboa,
Campo Grande, Edif\'icio C8, Lisboa 1749-016,  Portugal.
    \\
$^4$Faculty of Physics, University of Warsaw, ul. Pasteura 5,
PL--02--093 Warszawa, Poland.}

\vspace{10pt}
%
%\begin{indented}
%\item[]December 2019
%\end{indented}
\begin{abstract}
We describe possibilities of spontaneous, degenerate four-wave
mixing (FWM) processes in spin-orbit coupled Bose-Einstein
condensates. Phase matching conditions (i.e., energy and momentum
conservation laws) in such systems allow one to identify four
different configurations characterized by involvement of distinct
spinor states in which such a process can take place. We derived
these conditions from first principles and then illustrated dynamics
with direct numerical simulations. We found, among others, the
unique configuration, where both probe waves have smaller group
velocity than pump wave and proved numerically that it can be
observed experimentally under proper choice of the parameters. We
also reported the case when two different FWM processes can occur
simultaneously. The described resonant interactions of matter waves
is expected to play important role in the experiments of BEC with
artificial gauge fields. Beams created by FWM processes are
important source of correlated particles and can be used the
experiments testing quantum properties of atomic ensembles.
\end{abstract}

\vspace{2pc} \noindent{\it Keywords}: Spin-orbit coupling, four-wave
mixing, Bose-Einstein condensate.

\section{Introduction}
\label{sec:intro}

Traditionally four-wave mixing (FWM) processes are associated with
photon interactions via a non-linear polarization. It is a
third-order parametric process in which two particles (from two
so-called writing or pump beams - one particle from each beam) are
annihilated when passing through a non-linear medium, and at the
same time two new particles (constituting probe and signal beams)
are generated. In optics, FWM is commonly associated with the third
order Kerr nonlinearity. The phenomenon is ubiquitous (see e.g.,
~\cite{Agrawal, Boyd}) and its applications are very widespread,
including: fiber optic communication (very often not welcome),
wavelength conversion, parametric amplification, optical
regeneration, optic phase conjugation and correction of the
aberration of images.

FWM can be observed also for massive particles as it was predicted
and observed in cold atomic gasses two decades
ago~\cite{BECFWM}-\cite{BECatomlaser} (see also~\cite{Meystre}). In
this case resonantly interacting particles are neutral atoms rather
than photons. Flexibility of control of trapping potentials, as well
as of nonlinear interactions, in atomic systems open interesting
perspectives of managing both momentum conservation and energy
conservation laws through the interplay of additional linear and
nonlinear potentials. This issue was already explored in
Refs.~\cite{FWM_lin_latt1}-\cite{FWM_nonlin_latt2}. In this paper we
explore a similar idea of controlling  FWM processes through
artificially created gauge potentials.

In order to consider FWM in a specific medium, one has to identify
the characteristic {eigenmodes}: the elementary solutions to the
linearized equations of motion in the form of plane waves,
identified by their wavevectors and frequencies which satisfy the
so-called {phase matching conditions}, that are equivalent to
momentum and energy conservation laws. These are often quite
demanding constraints depending on the particular form of dispersion
relation characteristic for the system under investigation. For
instance, in one dimension they cannot be satisfied for a system of
cold atoms obeying parabolic dispersion relation and confined to
(quasi-)one dimension. The situation can be improved by artificial
change of the dispersion law using linear optical
lattices~\cite{FWM_lin_latt1,FWM_lin_latt2,FWM_lin_latt3} or by the
manipulation of the wavenumbers of the matter waves involved in the
process by means of nonlinear lattices
\cite{FWM_nonlin_latt1,FWM_nonlin_latt2}. These modifications
introduce the internal texture to the propagation medium, making it
inherently inhomogeneous.

If a system has a spinor nature, i.e., consists of two subsystems,
an alternative way to manipulate the linear properties of the
medium, even preserving homogeneity, is to employ coherent coupling
of the constituents. In optics, for example, one can satisfy the
matching conditions for the FWM of light propagating in homogeneous parity-time ($\mathcal{PT}$) - symmetric
coupled waveguides with gain and losses~\cite{FWM-PT}. One can also consider the matching conditions, and thus observation of the FWM in $\mathcal{PT}$-symmetric optical lattices  which are available experimentally~\cite{PT-LATTICE}. And yet another interesting application was demonstrated in multi-component vector solitons consisting of two perpendicular FWM dipole components created by electromagnetically induced gratings~\cite{FWM-GRAT}.

Similar situation naturally occurs  for spinor Bose-Einstein, where coupling between two atomic states by means of the spin-orbit coupling (SOC) allows one to manipulate the dispersion relation in the presence of external potential. This idea becomes attractive, since using various experimental techniques, spin-orbit-coupled Bose-Einstein condensates (SOC-BECs) of hyperfine states of ${}^{87}$Rb atoms has been created~\cite{SOC-Nature,SOC-2,SOC3,SOC-review1,SOC-review2,SOC-1}. Notice that atoms of s and p bands of the static lattice were considered as pseudospins~\cite{SOC4}.

The main goal of this paper is to show that with properly adjusted
SOC, one can satisfy the phase matching conditions for a homogeneous
one-dimensional SOC-BECs. Importantly, the SOC properties in atomic
systems are highly
adjustable~\cite{SOC-tuunable1,SOC-tuunable2,SOC-tuunable3,SOC-tuunable4}
and the matching conditions reported below are experimentally
feasible. Inter-atomic interactions are also tunable, most commonly
by Feshbach resonance; see Ref.~\cite{SOC2} for the observation of
Fesbach resonances for SOC fermions, or Ref.~\cite{SOC1} for
observation of partial waves, with nonlinear interactions controlled
by the dressing technique. Additional degrees of freedom in
manipulating the effective dispersion relation, and thus, of the
matching conditions [see (\ref{cons_moment}) and (\ref{cons_energy})
below], may be reached by using moving lattices~\cite{SOC3,SOC4}.

Understanding dynamics of FWM process is important in the context of creation od pairs of correlated particles, which are crucial in the ultra-precise metrology. Particles undergoing elastic collisions may create entanglement and it can be of practical use in the experiment. Here we mention the experiments~\cite{West1,West2,West3}, study of Cauchy Schwarz inequality~\cite{Schwarz1,Schwarz2}, and twin beam generation~\cite{Twin}. Also, recently there was a  prediction~\cite{Bell1} and its experimental confirmation~\cite{Bell2}, of the Bell inequality is violated in a many-body system of massive particles with spinor condensates.

In the context of present considerations we would like to mention
two recent experimental achievements. In the first a stripe phase
with supersolid properties in  SOC-BECs has been observed
\cite{SOC5}. In this case spin flip process with a momentum transfer
has been realized and observed using Bragg scattering. In another
experiment radio-frequency (rf) photons were dressed with tunable
recoil momentum by combining rf pulses with an oscillating magnetic
force. This leads to a new application of Floquet engineering:
periodically driven systems can have time-averaged properties which
cannot be achieved with constant fields and in our opinion hold a
promise of mixing different waves \cite{SOC6}.

Due to the spinor nature of the one-dimensional (1D) SOC-BEC, it is
characterized by two branches of the dispersion relation. As a
result, the matching conditions can be readily satisfied, as we
shall see below. Moreover, unlike in the case BECs without SOC,
where additional external potentials, like
linear~\cite{FWM_lin_latt1,FWM_lin_latt2,FWM_lin_latt3} or
nonlinear~\cite{FWM_nonlin_latt1,FWM_nonlin_latt2} lattices are
employed for ensuring matching conditions, now one can find  a {\em
diversity of distinct FWM processes}, where the interacting waves,
as well as waves generated may represent different spinor states at
the same values of parameters of the system (similarly to the FWM
with laser pulses reported in~\cite{FWM-PT}). It is a goal of the
present study to identify the FWM processes available in the FWM in
SOC-BECs and show they can be efficient enough to be observed.

The organization of the paper is as follows: in
Sec.~\ref{sec:FWM_BEC} we discuss phase matching conditions in a
quasi-1D SOC-BEC,  and we focus on the degenerate case, where the
central pump wave serves as a source for stimulated enhancement of
two probe waves. Here we borrow the terminology from optics but we
make no distinction between signal and probe beams. Next we identify
four possible configurations of FWM and in Sec.~\ref{sec:numerics}
we perform feasibility study using real time simulations for all
predicted configurations. The outcomes are summarized in Conclusion.

\section{Phase matching conditions}
\label{sec:FWM_BEC}
\subsection{\textit{\textbf{General relations}}}
Let us consider a quasi-1D SOC-BEC which is described by a
two-component order parameter
$\mathbf{\Psi}(x,t)=(\Psi_1(x,t),\Psi_2(x,t))^\mathrm{T}$ (hereafter
$\mathrm{T}$ stands for transposition). The dynamics of the spinor
$\mathbf{\Psi}(x,t)$ is governed by the coupled Gross-Pitaevskii
equations (GPEs):
\begin{eqnarray}
\label{GPE} i\partial_t \mathbf{\Psi} = H\mathbf{\Psi} +\frac{1}{2}
G(\mathbf{\Psi})\mathbf{\Psi},
\end{eqnarray}
where
\begin{eqnarray}
\label{hamiltonian}
H=\frac{1}{2}\left(-\partial_x^2+\mathbf{\Omega}\cdot
\mathbf{\sigma} -i\alpha\sigma_x\partial_x\right)
\end{eqnarray}
is the linear mean field Hamiltonian of the two component BEC,
$\alpha$ is the SOC strength, $\mathbf{\Omega}$ is the vector of the
Zeeman coupling (we admit external magnetic field),
$\mathbf{\sigma}=(\sigma_x,\sigma_y,\sigma_z)$ is the vector of
Pauli matrices $\sigma_{x,y,z}$. Physical interpretation of these
parameters depends on the particular realization. For instance in
the experiment performed in an $^{87}$Rb Bose-Einstein condensate, a
pair of Raman lasers created a momentum sensitive coupling between
two internal atomic states \cite{SOC-Nature}. This SOC was
equivalent to that of an electronic system with equal contributions
of Rashba and Dresselhaus couplings, and with a uniform magnetic
field $B$ in the $(x,z)-$plane. In materials the
SOC is due to intrinsic properties, which are largely determined by
the specific material and the details of its growth. In these and
other proposed schemes $\mathbf{\Omega}$ may have different physical
interpretation, including Rabi frequencies of the dressing laser
fields. It has to be taken into account when one consider tuning and
range of vector $\mathbf{\Omega}$. Here we will call it Zeeman
coupling, and $\Omega_z$ Zeeman splitting.

The nonlinearity $2\times 2$ matrix is given by
\begin{equation}
G(\mathbf{\Psi})=\left(
\begin{array}{c c}
g_1|\Psi_1|^2+g|\Psi_2|^2 & 0\\
0 & g|\Psi_1|^2+g_2|\Psi_2|^2\\
\end{array} \right) \label{nonlinearity1}
\end{equation}
with intra-- and inter--component interactions $g_{1,2}$ and $g$,
respectively, and we use the dimensionless units with $\hbar=m=1$.

To address the matching conditions for the FWM we start with the
eigenmodes of the linear spectral problem, representing them in the
form of the plane waves
\begin{equation}
\label{psi}
\mathbf{\Psi}_{\pm}(x,t) = e^{i k x-i \mu_\pm(k)t}
\mathbf{\psi}_\pm(k),
\end{equation}
where
$\mathbf{\psi}_\pm(k)=(\psi_\pm^{(1)}(k),\psi_\pm^{(2)}(k))^{\rm T}$
is a constant (i.e. $x$- and $t$-independent) spinor, $k$ is a mode
wavenumber, $\mu_\pm(k)$ is its frequency, and $\pm$ indicate the
upper ("$+$") and lower ("$-$") branches of the spectrum [i.e.,
$\mu_-(k)\leq \mu_+(k)$]. We will concentrate in the case of the
magnetic field in the $(x,z)$ plane, i.e.
$\mathbf{\Omega}=(\Omega_x,0,\Omega_z)$ [Without loss of generality
we fix $\Omega_{x},\Omega_{z}\geq 0$], for which  we compute the two
branches of the dispersion relation
 \begin{equation}
 \label{eq:chempot}
 \mu_{\pm}(k)=\frac{k^2}{2} \pm \frac{\varepsilon(k)}{2}.
 \end{equation}
Here
\begin{equation}
\varepsilon (k)= \sqrt{\Omega_z^2+\widetilde{\Omega}^2(k)}
\end{equation}
with $\widetilde{\Omega}(k)=\alpha k+\Omega_x$, is the gap between
the spectral branches at a given $k$. The lower (-) and the upper
(+) branches of the respective eigenvectors are defined as
    \begin{equation}
     \label{eigenvector}
    \mathbf{\psi}_\pm(k)=\frac{1}{\sqrt{\widetilde{\Omega}^2+(\varepsilon(k)\mp \Omega_z)^2}}
    \left(\begin{array}{c}
    \widetilde{\Omega}
    \\
    \pm \varepsilon(k)-\Omega_z
    \end{array}\right).
    \end{equation}

  It is to be emphasized that (\ref{psi}), as well as the matching conditions (\ref{cons_moment}), (\ref{cons_energy}), strictly speaking are valid only in linear regime, or approximately valid during the initial stages of the evolution. If however, the pulses with newly generated central frequencies separate in space from the initial pulse due to different group velocities, further generation of new harmonics is suppressed and one can observe propagation of quasi-monochromatic pulses. Confirmation of the described scenario requires numerical simulation of the full nonlinear problem. This approach is adopted below.

To reduce the number of parameters, here  we investigate the
degenerate FWM process, where  two input spinor states are
identical. We label their wavevectors by $k_1$ and using the optical
terminology, we call them pump waves. Two spinors that are created
in the FWM process with central wavevectors $k_2$ and $k_3$, will be
referred to as probe waves. Respectively,  the conservation of
wavenumbers and frequencies of the pump and probe waves, are
expressed in the from of phase-matching conditions
\begin{eqnarray}
\label{cons_moment}
&2k_1= k_2 + k_3\,,\\
\label{cons_energy}
&2\mu_{\nu_1}(k_1)=\mu_{\nu_2}(k_2)+\mu_{\nu_3}(k_3).
\end{eqnarray}
{The indexes} $\nu_j$ ($j=1,2,3$) refer to either ``$+$'' or ``$-$''
{branch of the spectrum}.

Below we consider only the cases $k_{2,3}\neq k_1$, excluding the
trivial case of self-phase modulation where all wavenumbers are
equal. We note that the system (\ref{GPE})  obeys gauge, rather than
Gallilean, invariance, at which the generalized momentum
\begin{eqnarray}
\Pi=\int_{-\infty}^{\infty}\mathbf{\Psi}^\dagger\left(-i\partial_x+\frac{\alpha}{2}\sigma_x\right)\mathbf{\Psi}
dx
\end{eqnarray}
is conserved: $d\Pi/dt=0$. For our consideration this means that the
input wavenumber $k_1$ cannot be set arbitrarily to zero without
changing the spinor eigenstates. It also means that at zero Zeeman
field $\mathbf{\Omega}=0$, the linear Hamiltonian $H$ is gauge
equivalent to the usual one-dimensional Schr\"odinger Hamiltonian
$H_0=-\partial_x^2$ which does not support the matching conditions
(\ref{cons_moment}) and (\ref{cons_energy}). In other words, while
SOC controls the waves involved in resonant processes, the FWM
itself requires nonzero Zeeman field.

For the following consideration it is convenient to rewrite
(\ref{cons_moment}) as
\begin{equation}
\label{cons_moment_1} k_2 = k_1 + q ,\quad k_3 = k_1 - q.
\end{equation}
Now matching condition for frequencies (\ref{cons_energy}) can be
rewritten in the form
\begin{equation}
2q^2=2s_1\varepsilon(k_1)-s_2\varepsilon\left(k_1+q\right) -
s_3\varepsilon\left(k_1-q\right), \label{eq:ph_matching_cond}
\end{equation}
where $s_j=\pm1$. Since each wave belongs to either upper or lower
branch, these are eight different equations for given $k_1$ and $q$.
However only four of them have nontrivial solutions. To justify this
we first notice that if $s_2=s_3$, the condition
(\ref{eq:ph_matching_cond}) is symmetric under $q\leftrightarrow -q$
exchange. If however $s_2\neq s_3$, then (\ref{eq:ph_matching_cond})
is symmetric with respect to simultaneous change
$(s_2,q)\leftrightarrow(s_3,-q)$. This allows one to restrict the
analysis to the case $q>0$. Next we use $\varepsilon(k)>0$ property
and conclude that the case $(s_1,s_2,s_3)=(-1,1,1)$ does not have
solutions since the right hand side of
Eq.~(\ref{eq:ph_matching_cond}) becomes negative. Let us now
consider $k_1\geq 0$ (the case $k_1<0$ is fully analogous). For
non-negative values of $k_1$ we find the following inequalities
\begin{eqnarray}
\label{ineqs} 0 \leq  q^2
  \leq\varepsilon_+(k_1-q)\leq \varepsilon_+(k_1+q),
\end{eqnarray}
excluding the cases $(-1,1,-1)$ and $(-1,-1,1)$. Hence, the initial
pulse from the lower brunch of the spectrum may originate degenerate
FWM in processes involving modes from the lowest branch only (this
is the configuration 4 in the Table~\ref{tab:approx} below).
Finally, using inequalities (\ref{ineqs})  one can exclude also the
case $(1,1,1)$, leaving only four possible configurations summarized
in the Table.~\ref{tab:approx}.
\begin{table}[h]
    \caption{Possible configurations of degenerate FWM processes (positive and negative $q$ are included). The first column are numbers identifying configurations, which corresponds to the specific choice of the spectrum branches for the pump and probe waves, indicated in second, third, and fourth columns, respectively. In the last column we show  the maximal number, $N_{\rm max }$, of $q$ values  solving Eq.~(\ref{eq:ph_matching_cond}).}

\centering

        \begin{tabular}{|c|c|c|c|c|}
        \hline
        Configuration & $s_1$ & $s_2$ & $s_3$ & $N_{\rm max}$\\
        \hline\hline
        1 & $1$ & $-1$ & $-1$ & 2\\
        \hline
        2 & $1$ & $1$ & $-1$ & 2\\
        \hline
        3 & $1$ & $-1$ & $1$ & 2\\
        \hline
        4 & $-1$ & $-1$ & $-1$ & 4\\
        \hline\hline
    \end{tabular}

\label{tab:approx}
\end{table}

In the last column of the Table.~\ref{tab:approx} we list the
maximal number of solutions for particular configuration. In what
follows we present analytical and graphic considerations that led us
to these counts.

\subsection{\textit{\textbf{Analysis of possible configurations}}}

By straightforward algebraic manipulations we can eliminate
square-root terms in the equation (\ref{eq:ph_matching_cond})
(simple sequence of transfers and squaring).  As a result all four
phase matched processes listed in the Table \ref{tab:approx} are
determined by the following cubic equations
%\begin{widetext}
%    \begin{equation}
%    \label{eq:ss}
%      Q^3-\left(1+4s_1\sqrt {{\tomega}^{2}+\omega_z^{2}}\right)Q^2+\left(2s_1\sqrt {{\tomega}^{2}+\omega_z^{2}} +5{\tomega}^{2}+5 \omega_z^{2} \right)Q
%     -\omega_z^2-2s_1\sqrt{\tomega^2+\omega_z^2}\left(\tomega^2+\omega_z^2\right)=0
%    \end{equation}
%    \end{widetext}
\begin{eqnarray}
\label{eq:ss}
Q^3-\left(1+4s_1\sqrt{{\tomega}^{2}+\omega_z^{2}}\right)Q^2&
+\left(2s_1\sqrt{{\tomega}^{2}+\omega_z^{2}}
+5{\tomega}^{2}+5\omega_z^{2}\right)Q  \nonumber
\\
&-\omega_z^2-2s_1\sqrt{\tomega^2+\omega_z^2}\left(\tomega^2+\omega_z^2\right)=0
\end{eqnarray}
where $\widetilde{\omega}= \widetilde{\Omega}/\alpha^2$,
$\omega_z=\Omega_z/\alpha^2$ and $Q=q^2/\alpha^2$. For obvious
reasons we are interested only in positive roots of (\ref{eq:ss})
and exclude the root $Q=0$ (i.e. $q=0$) which does not correspond to
FWM but to the self-phase modulation.

A number of real roots  of Eq.~(\ref{eq:ss}) is determined by the
sign  of the discriminant
\begin{eqnarray}
%\begin{equation}
 \label{eq:dsc}
\Delta_{s_1}={\tomega}^{2} \left[15\tomega^2  +4s_1\sqrt
{{\tomega}^{2}+\omega_z^{2}}\left({\tomega}^{2}+ \omega_z^{2}+3
\right)-12\omega_z^{2}-4\right].
%\end{equation}
\end{eqnarray}
If $\Delta_{s_1}>0$, there exists one real root. Three distinct real
solutions exist if $\Delta_{s_1}<0$. At $\Delta_{s_1}=0$ all roots
are real and at least one is multiple \cite{Vieta1a}.

Now we inspect systematically configurations listed in
Tab.~\ref{tab:approx} which manifest qualitatively different types
of dynamics. Starting with the last one, we set $s_1=-1$. Now the
discriminant (\ref{eq:dsc}) depends on the two parameters
$\{\widetilde{\omega},\omega_z\}$ for different values of which
$\Delta_{s_1}$ can acquire any sign or be zero. Analyzing Vieta
formulae one can exclude the possibility of all three real roots
being positive \cite{Vieta1a}. It means that in the
configuration 4 (see the Table~\ref{tab:approx}) there may exist
either one or two real positive roots of Eq.~(\ref{eq:ss}).  Taking
into account that roots appear in pairs, $\pm q$, this corresponds
to at most four possible arrangements allowed by the phase matching
condition (\ref{eq:ph_matching_cond}).

Next we turn to the configurations 1, 2, and 3 in the
Table~\ref{tab:approx}, by setting $s_1=+1$ in (\ref{eq:ss})  and in
(\ref{eq:dsc}).  In this case we find that $\Delta_{s_1}<0$  for all
values of the parameters $\{\widetilde{\omega},\omega_z\}$, i.e. all
roots are real. Moreover, they are all positive as follows from the
Vieta formulae. The three real positive roots of $Q$, i.e. six roots
$q$, correspond to the three different configurations [notice that
the difference among these configurations was removed upon squaring
in obtaining Eq.~(\ref{eq:ss})].

It is easy to  establish one-to-one correspondence of the roots and
the configurations. For if one root is common for two configurations
we use equation (\ref{eq:ph_matching_cond}) to show that one of the
terms $\varepsilon_s(k_1\pm q)$ is equal to zero. Obviously, for
each of the configurations with $s_1=+1$ always there is at least
one positive root (all roots cannot be negative, because their
product is positive). Let $0< q_{1},q_2,q_3$ denote the  positive
roots, so that the configuration 1 has two symmetric roots, which we
denote by $\pm q_1$ while the roots of the configurations 2 and 3
are given by ($q_2$, $-q_3$) and ($-q_2$, $q_3$), respectively.

\begin{figure}[h]
    \centering
     \includegraphics[width=0.90\columnwidth]{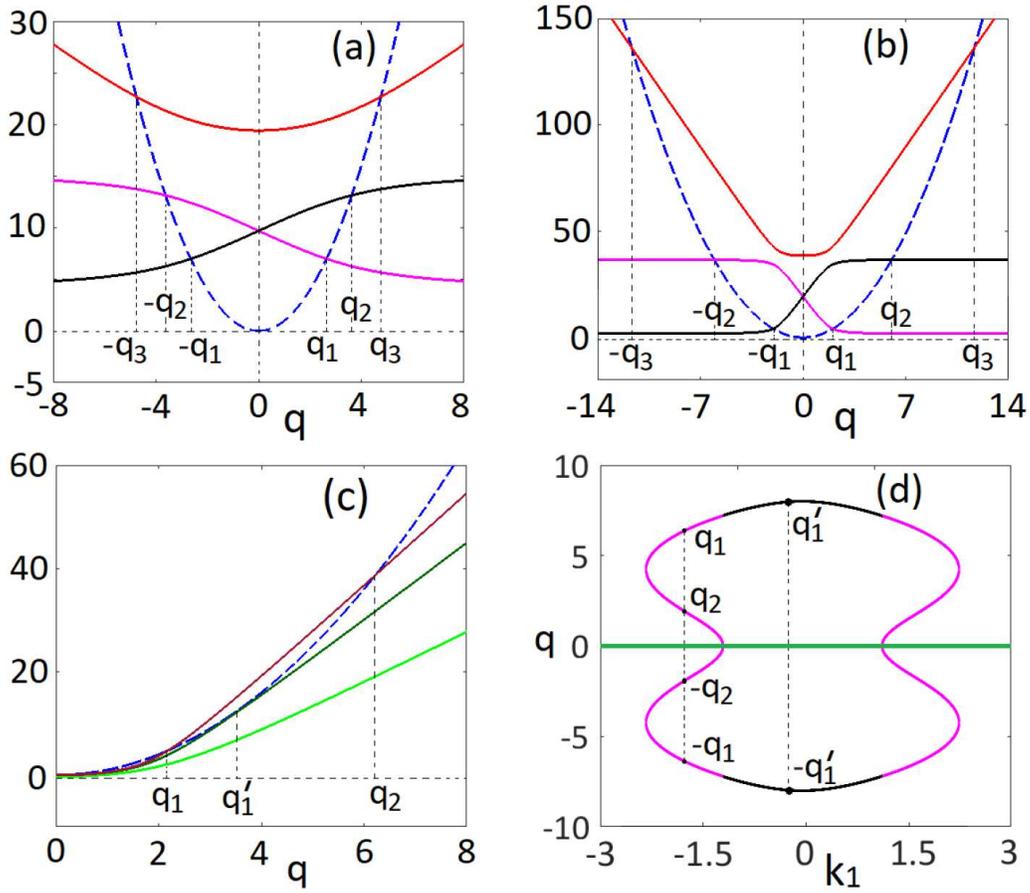}
    \caption{
        The graphic representation of phase matching equation. Panels (a), (b) and (c) where LHS (blue dashed curve) and RHS (solid curves) of Eq.~(\ref{eq:ph_matching_cond}) are presented as a function of parameter $q$, for $\Omega_x=2.5$, $\Omega_z=8$,
        and $k_1=1.5$. The roots $\pm q_{1,2,3}$ of Eq.~(\ref{eq:ph_matching_cond}) are located at the crossing of the blue dashed curve and solid curves.
        In (a) and (b) the SOC strength is $\alpha=2$ and $\alpha=10$, respectively. The red, pink and black lines correspond to the configurations 1, 2 and 3 (see table (\ref{tab:approx})).
        Panel (c) exemplifies configuration 4; here light green line represents RHS of Eq.~(\ref{eq:ph_matching_cond}) for $\alpha=5$, dark blue for $\alpha=7.67$  and brown for $\alpha=9$.  In panel (d) we fix  $\alpha=9$ and vary $k_1$ to show regions of zero, one and two roots.
        }
    \label{fig:123_cases}
 \end{figure}

These arguments are exemplified in Fig.~\ref{fig:123_cases} where we
show graphical solutions of phase matching equation. In the panels
(a), obtained for $\alpha=2$, and (b), obtained for $\alpha=10$, the
dashed blue curve is the plot of the left hand side (LHS) of
Eq.~(\ref{eq:ph_matching_cond}). Right hand side (RHS) of this
equation is represented by red, pink and black lines correspond to
configurations 1, 2 and 3 in Tab.~\ref{tab:approx}. Crossings of
solid and dashed lines yield the roots of the phase matching
equations. Comparing the values of the LHS and RHS of
Eq.~(\ref{eq:ph_matching_cond}) at $q=0$ and at $q\to\infty$, we
conclude that each of the configurations 1, 2 and 3 must have at
least one root for $q>0$.
The fourth configuration ($s_1=s_2=s_3=-1$) is illustrated in panel
(c). Here we observe three different possibilities: no positive
solutions of  the phase matching condition
(\ref{eq:ph_matching_cond}) (the light green curve does not cross
the dash blue curve at $\alpha=5$); one positive root (the deep
green and dash blue curves tangent to each other when $\alpha\approx
7.67$), and two positive solutions (the brown curve crosses the dash
blue curve in two points, when $\alpha=9$).

Finally, in the panel (d) we varied $k_1$ (while holding $\alpha=9$)
and shown the regions where the phase matching equation of the
fourth configuration supports one positive (the black curves) and
two positive solutions (the pink curves). In all panels (a)-(d) we
fixed $\Omega_x=2.5$ and $\Omega_z=8$.

\subsection{\textbf{\textit{Matching of group velocities}}}

While no matching conditions on the group velocities  is imposed,
for practical observation of different scenarios of FWM in numerical
simulations, the issue of the group velocities (GVs)
\begin{eqnarray}
\label{def:GV} v_\pm(k) \equiv \frac{\partial \mu_\pm(k)}{\partial
k} = k \pm
\frac{\widetilde{\Omega}}{2\sqrt{\widetilde{\Omega}^2+\Omega_z^2}}\,,
\end{eqnarray}
becomes relevant. On the one hand where all wavepackets move with
respect to each other, it is important that the spinor involved in
the process have similar values of GVs: otherwise fast separation of
wavepackets in space may drastically reduce the conversion
efficiency. On the other hand, GVs should have sufficient difference
in order to observe spatial separation of the probe wavepackets.
Thus in addition to solving the matching conditions we set a task of
finding optimal conditions in the context of FWM numerical
simulations (they are presented in the next section).

For each configuration listed in
Table~\ref{tab:approx}, at a given $k_1$ one can determine $q$,
i.e.,  the wavenumbers $k_2$ and $k_3$, and consequently their GVs.

\begin{figure}[ht]
    \centering
    \includegraphics[width=0.99\columnwidth]{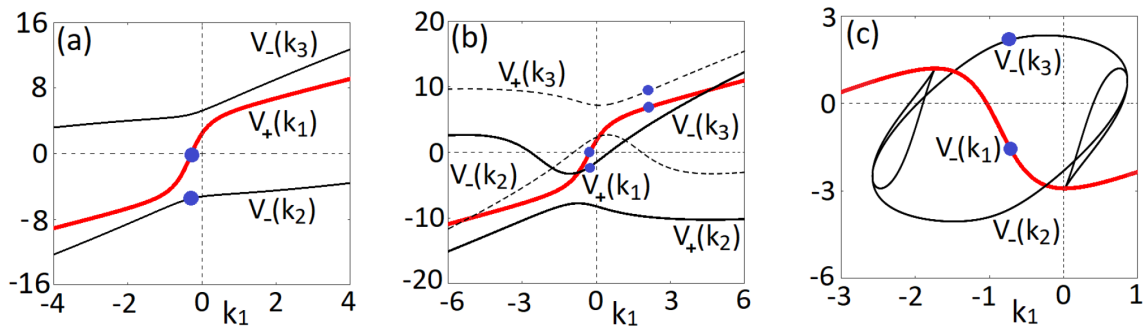}
    \caption{The GVs associated with different
        configurations of degenerated FWM presented in
        Table~\ref{tab:approx} versus momentum of the pump wave. Panel (a) shows the first configuration with
        $\Omega_x=2.5$, $\Omega_z=4$ and $\alpha=3$. Panel (b) shows second and third configurations, with $\Omega_x=3$,
        $\Omega_z=8$ and $\alpha=10$, by the solid and dashed black lines, respectively. Panel (c) illustrates the forth configuration with
        $\Omega_x=6$, $\Omega_z=4$ and $\alpha=7$. In all panels, the group
        velocities of the pump (probe) wavepackets are shown by thick red
        (black) curves. The blue dots marked on the red and black curves indicate the points where we do numerical simulations. The dynamics of FWM at these particular points are shown in the Figures \ref{fig:first_fwm_process} - \ref{fig:forth_fwm_process}.} \label{fig:set of GVs}
\end{figure}
%At first, we note that the GVs of different branches cross each
%other at $k_0=-\Omega_x/\alpha$, when $\widetilde{\Omega}(k_0)=0$.
%Away
%from the crossing point we have two parallel linear asymptotes:
%$v_\pm(k)\to k\pm 1/2 $ at $k\to\infty$.
In Fig.~\ref{fig:set of
GVs} (a)  we observe that the GV of the pump wavepacket, $v_+(k_1)$,
is close to either $v_-(k_2)$ or $v_-(k_3)$, almost for all $k_1$
except the vicinity of $k_1=0$. This means that to obtain clear
separation of the wavepackets, generated in the FWM at relatively
short time intervals, $k_1$ should be chosen close to zero. Then,
the separation between velocities grows rapidly enough allowing
direct observation of separated pulses. On the other hand, the time
that pulses overlap is still long enough to generate a substantial
four wave mixing signal.
\begin{figure}[th]
    \centering
        \includegraphics[width=0.90\columnwidth]{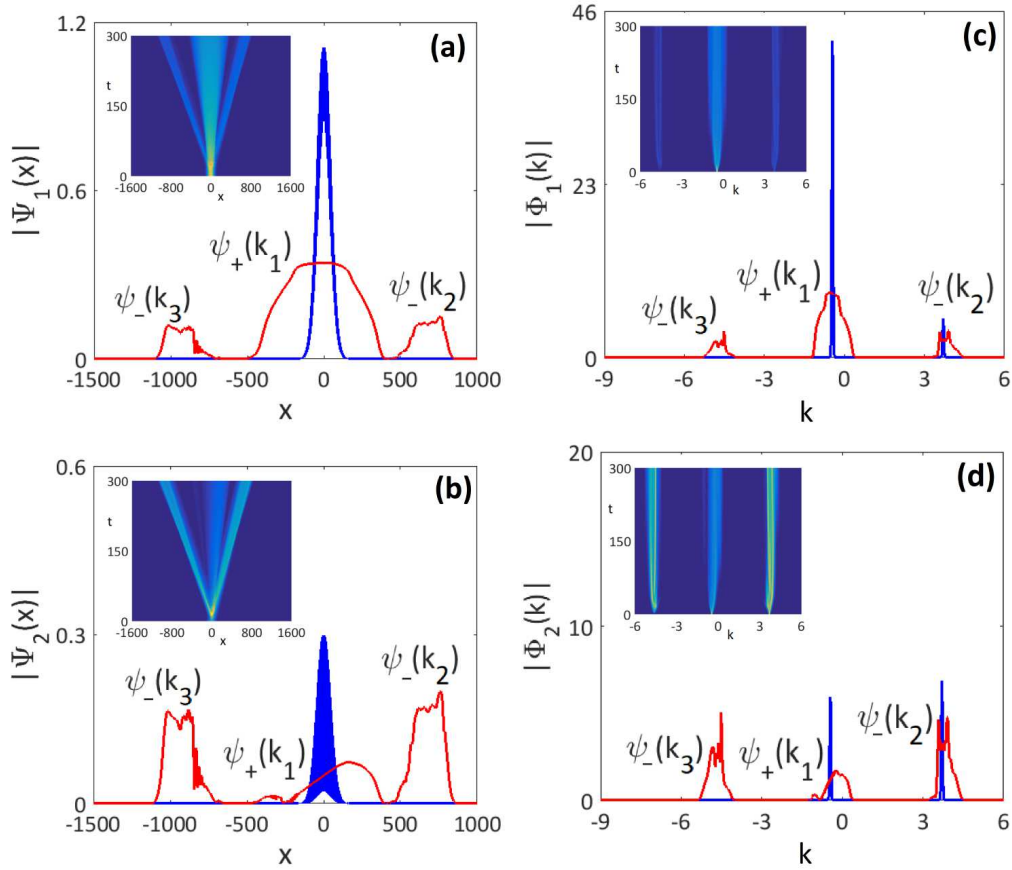}
    \caption{Initial (blue) and final (red) states of FWM process of configuration 1 from
        Table~\ref{tab:approx}. The moduli of the first component $|\Psi_1(x,t)|$
        and of its Fourier transform $|\Phi_1(k,t)|$ are shown in upper
        row while the corresponding quantities of the second component are
        presented in the lower row. Insets show the respective
        temporal evolutions. The parameters are: $\Omega_z=4$, $\alpha=3$,
        $\Omega_x=2.5$, $k_1=-0.45$, $k_2=3.704$, $k_3=-4.604$, $g=0.8$,
        $g_1=0.808$, $g_2=0.792$. Time of the evolution is equal to $t=300$, the total norm $N\approx 78$, $A_1=1$, $A_2=0.2$, $A_3=0$ and $w=60$. The correspondence between the picks and the spinor states is indicated inside each panel.}
    \label{fig:first_fwm_process}
\end{figure}

Fig.~\ref{fig:set of GVs} (b) shows the dependence of GVs of phase
matched wavepackets versus pump momentum for configurations 2 and 3
from the Table~\ref{tab:approx}.  We again observe that in some
regions GVs are close to each other or even coincide what does not
allow observation of separation of the generated wavepackets from
the initial one. However an interesting situation occurs in the
vicinity of $k_1=0$. Here GVs of the second and third waves have
bigger absolute values than $v_+(k_1)$ and the same sign. In this
case both created waves move faster that the initial wavepacket. We
should emphasize that this is not common in the usual realizations
of FWM and this is solely due to the SOC coupling. Note that in
these configurations it is also possible to initiate FWM process
where one of the velocities of the probe waves is smaller and one
bigger than that of the pump.
\begin{figure}[th]
    \centering
    \includegraphics[width=0.90\columnwidth]{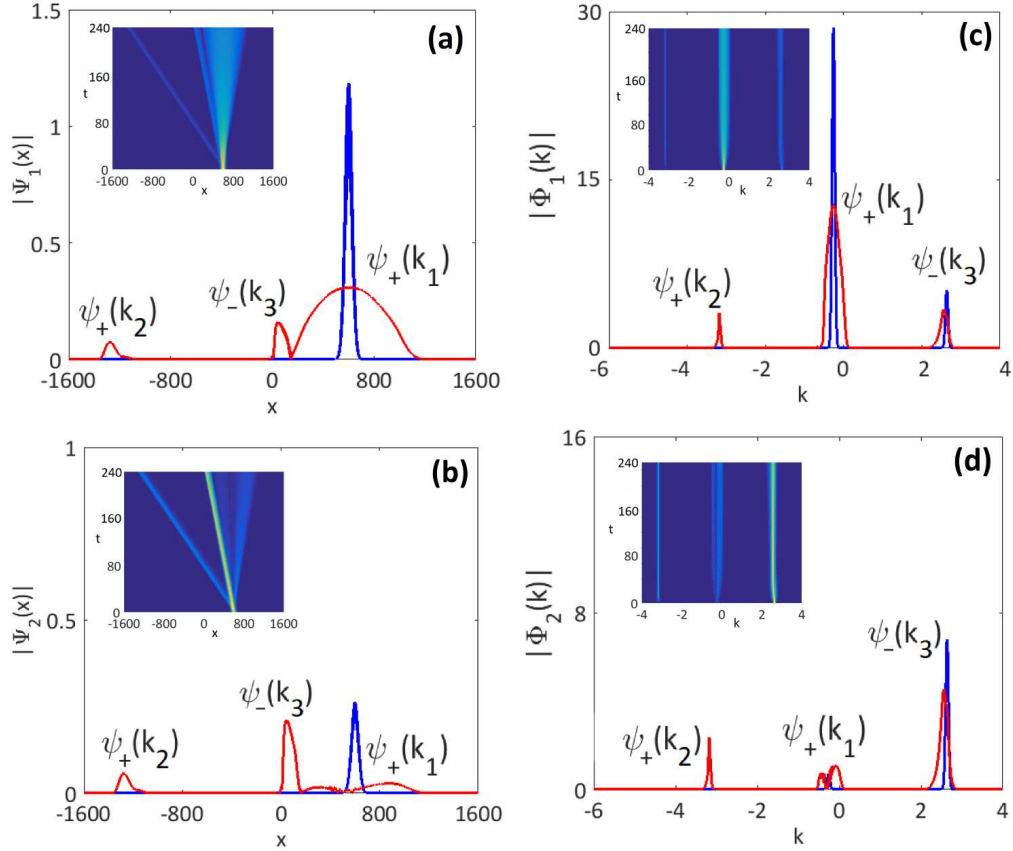}
    \caption{Initial (blue) and final (red) states of FWM process of the second configuration from Table~\ref{tab:approx}. The amplitudes of the first spinor component $|\Psi_1(x,t)|$
and of its Fourier transform $|\Phi_1(k,t)|$ are shown in upper row
while the corresponding quantities of the second spinor component
are presented in the lower row. Insets show the respective temporal
evolutions.   Here we illustrate the second FWM process with values
of parameters: $\Omega_z=8$, $\alpha=10$, $\Omega_x=3$, $k_1=-0.26$,
$k_2=-3.158$, $k_3=2.638$, $g=0.3$, $g_1=0.303$, $g_2=0.297$. Time
of evolution $t=240$ and the total norm $N\approx 56$. Initial vales
of amplitudes are $A_1=1$, $A_2=0$, $A_3=0.3$ and the width $w=40$.
The correspondence between the picks and the spinor states is
indicated inside each panel. }\label{fig:second_fwm_process}
\end{figure}
The situation is more complicated for the case of the fourth
configuration in Table \ref{tab:approx}, as shown in
Fig.~\ref{fig:set of GVs} (c). The (thick) red curve represents GV
of the pump wave of the negative branch $v_-(k_1)$ [see
Eq.~(\ref{def:GV})]. The other (black) curves, that have forms of
three loops, represent GVs of generated waves [$v_-(k_2)$ and
$v_-(k_3)$] that correspond to other (non-trivial) solutions. Like
in the previous cases, to reach significant separation of the pulses
in the real space we choose $k_1$ in a region far from the crossing
of the curves.

\section{Numerical results}
\label{sec:numerics}

Equipped with the solutions of matching conditions and with the
ideas of optimization the conversion efficiency in terms of the GVs
we now turn to direct numerical simulations of the configurations of
the FWM processes summarized in Table~\ref{tab:approx}. In order to
find favorable conditions to observe clear evidence of specific FWM
process, first one has to select proper momentum $k_1$. Note that
phase matching will automatically determine all participating
wavepackets GVs as explained in the previous section. Then,
appropriate initial widths and amplitudes of the pump and probe
waves need to be adjusted to ensure long enough and strong enough
nonlinear interaction.

In all simulations we use the wavepackets having equal widths and
completely overlapping at $t=0$, i.e.,
\begin{eqnarray}
\mathbf{\Psi}(x,t=0)= e^{-x^2/w^2}\sum_{j=1}^3 A_j
\mathbf{\psi}_{s_j}(k_j) e^{ik_jx}. \label{eq:initial}
\end{eqnarray}
Here\ $A_j$ are initial amplitudes of the wave-packets with the
central wavevectors $k_j$ of the spinors defined in accordance with
Eq.~(\ref{eigenvector}).
\begin{figure}[th]
\centering
    \includegraphics[width=0.90\columnwidth]{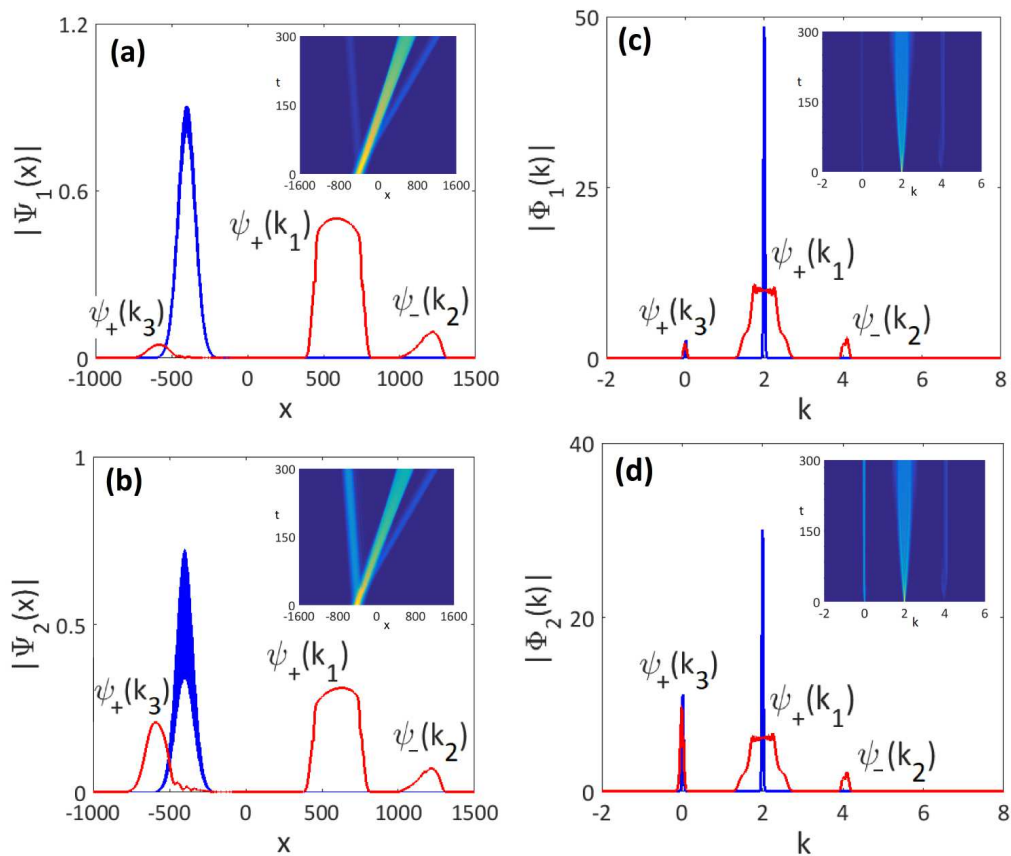}
    \caption{Initial (blue) and final (red) states of FWM process of the third configuration from Table~\ref{tab:approx}. The moduli of the first spinor component $|\Psi_1(x,t)|$
and of its Fourier transform $|\Phi_1(k,t)|$ are shown in upper row
while the corresponding quantities of the second spinor component
are presented in the lower row. Insets show the respective temporal
evolutions. Values of other parameters are:
        $\Omega_z=4$, $\alpha=3$, $\Omega_x=2$, $k_1=2$,$k_2=3.984$, $k_3=0.0164$, $g=0.3$, $g_1=0.303$, $g_2=0.297$. Time of evolution
        $t=300$ and the total norm $N\approx 107$. Initial amplitudes are $A_1=1$, $A_2=0$, $A_3=0.2$ and width $w=80$.
    The correspondence between the picks and the spinor states is indicated inside each panel.
}\label{fig:third_fwm_process}
\end{figure}

\begin{figure}[th]
    \centering
    \includegraphics[width=0.90\columnwidth]{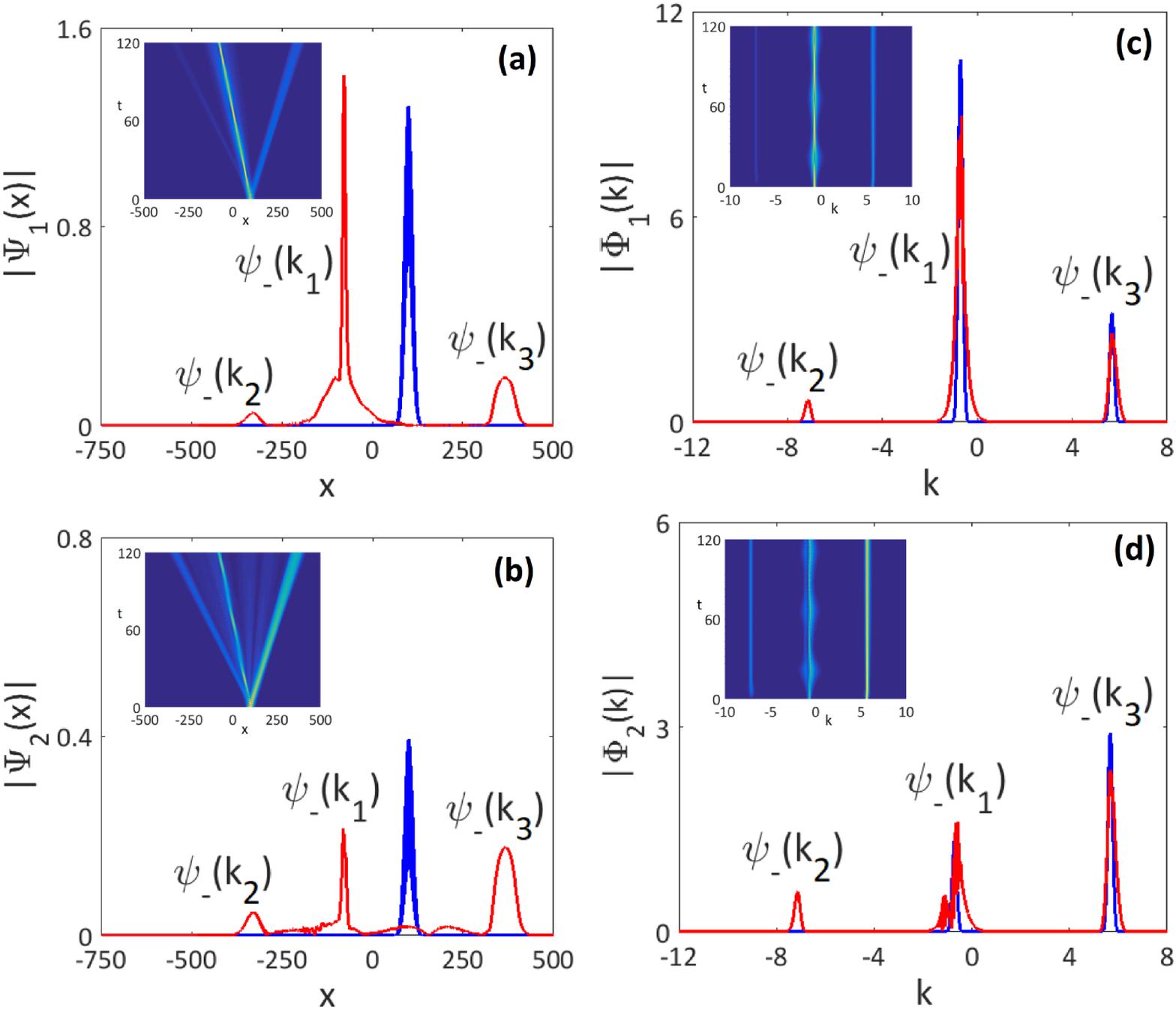}
    \caption{Initial (blue) and final (red) states of FWM process of the forth configuration from
Table~\ref{tab:approx}. The first spinor component $|\Psi_1(x,t)|$
and its Fourier transform $|\Phi_1(k,t)|$ are shown in upper row
while the corresponding quantities of the second spinor component is
presented in the lower row. Insets show the respective temporal
evolutions. Values of parameters:
        $\Omega_z=4$, $\alpha=7$, $\Omega_x=6$, $k_1=-0.71$, $k_2=-7.091$,
        $k_3=5.671$, $g=0.3$, $g_1=0.303$, $g_2=0.297$. Time of evolution
        $t=120$ and the total norm $N\approx 10$. Initial amplitudes are: $A_1=1$, $A_2=0$, $A_3=0.4$ and width $w=15$. The correspondence between the picks and the spinor states is indicated inside each panel.
    }\label{fig:forth_fwm_process}
\end{figure}

\begin{figure}[th]
    \centering
    \includegraphics[width=0.90\columnwidth]{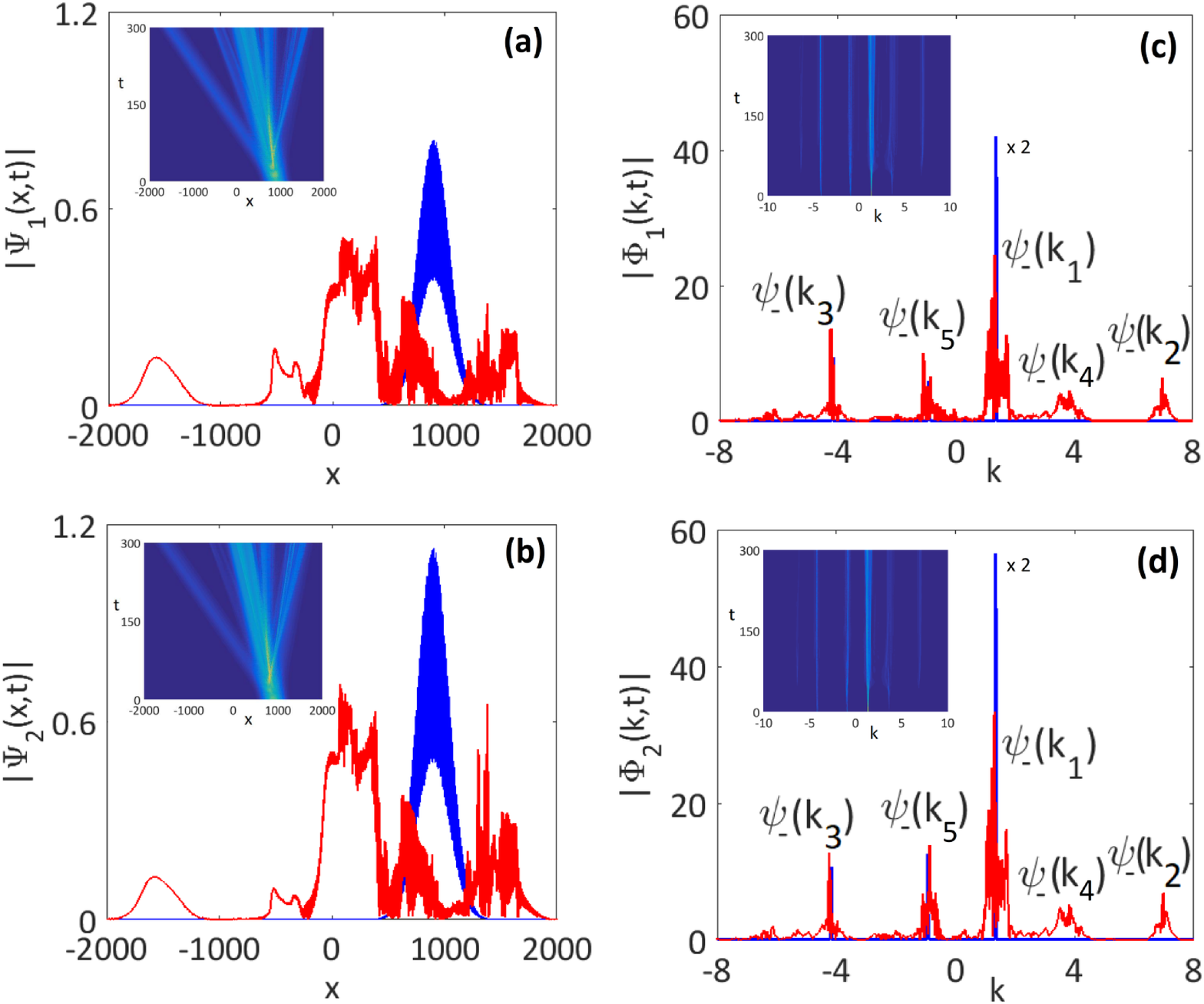}
    \caption{Initial (blue) and final (red) states of FWM process of the forth configuration from
Table~\ref{tab:approx}. The first spinor component $|\Psi_1(x,t)|$
and its Fourier transform $|\Phi_1(k,t)|$ are shown in upper row
while the corresponding quantities of the second spinor component is
presented in the lower row. Insets show the respective temporal
evolutions. Values of parameters:
        $\Omega_z=4$, $\alpha=8$, $\Omega_x=2.5$, $k_1=1.35$, $k_2=6.87$,
        $k_3=-4.17$, $k_4=3.635$, $k_5=-0.935$, $g=0.5$, $g_1=0.505$, $g_2=0.495$. Time of evolution
        $t=300$ and the total norm $N\approx 271$. Initial amplitudes are $A_1=1$, $A_2=A_4=0$, $A_3=A_5=0.2$ and width $w=200$. In the second and the fourth panels, we divided the initial Fourier components of spinors by factor 2, to improve the visibility of modes $k_2$ and $k_4$, created in the FWM process.
The correspondence between the picks and the spinor states is
indicated inside each panel. }\label{fig:forth_fwm_process4modes}
\end{figure}

For the FWM process corresponding to configuration 1, the initial
state is formed with $A_1=1$, $A_2=0.2$, $A_3=0$ and $s_j$ are
chosen according to Table~\ref{tab:approx}: $s_1=1$, $s_2=-1$ and
$s_3=-1$. In Fig.~\ref{fig:first_fwm_process} we show an example of
the FWM for this configuration. Due to the FWM process, by the
expense of the highly populated initial state $A_1$ we observe
strong amplification of the seed state and growth of the third
matter wave with phase matched momentum $k_3$.

This process is depicted with snapshots at the beginning and end
(i.e., at $t=0$ and $t=300$) of the simulations in
Fig.~\ref{fig:first_fwm_process}, where main panels (a), (c) [(b),
(d)] refer to the first [second] spinor component. In particular
blue contours in panels (a), (b) represent initially overlapping
pump ($k_1$) and probe ($k_2$) waves in the configuration space.
They are fully separated after evolution time ($t=300$) due to the
difference in GVs and new, clearly visible, wave of central momentum
$k_3$ is generated. Panels (c), (d) show the corresponding features
in the Fourier space. Here we distinct two waves as narrow blue
peaks, at initial time and again three waves at the end of
evolution. The most explicit feature is substantial broadening of
all participating matter waves during the evolution. The inset in
each panel shows full time evolution of  modulus of the spinor
components - (a) and (b) in the real space and (c) and (d) in the
momentum space.

In the next two figures we illustrate the FWM process corresponding
the second and third configuration from Table~\ref{tab:approx}, with
GV configuration shown in panel (b) of the Fig. \ref{fig:set of
GVs}. As mentioned above in these two configurations the exist  two
different roots ($q_2 \neq q_3$) of phase matching
condition~(\ref{eq:ph_matching_cond}). As one can see directly in
panels (a) and (b) of Fig.~\ref{fig:123_cases} these configurations
are related by the transformation $q_1 \rightarrow -q_2$ and $q_2
\rightarrow -q_1$, i.e. the analysis of second and third
configuration are analogous.

Interestingly, in Fig.~\ref{fig:second_fwm_process} both probe and
created waves are generated in the same side of the pump wave. The
evolution of the probe wavepackets in
Fig.~\ref{fig:third_fwm_process} looks qualitatively similar to that
one shown in Fig.~\ref{fig:first_fwm_process}. However, since each
newborn wavepacket bears a quasi-spin, the emergent spinors (more
precisely the left propagating waves) are different in these cases.

Turning to the fourth in the Table (\ref{tab:approx}) , we recall
that in this case one can obtain up to five solutions from the phase
matching condition (including the trivial case of $q=0$). We start
with Fig.~\ref{fig:forth_fwm_process}, where initial group
velocities can be identified in the panel (c) of the
Fig.~\ref{fig:set of GVs} and are marked as dots on red and black
curves. In principle, the dynamics presented in this case is very
similar to that shown for the configurations 1 and 3, except that
now different spinor states are involved (respectively the
wavepackets bear different quasi-spins). Also closer look at
Fig.~\ref{fig:forth_fwm_process} (c) and
Fig.~\ref{fig:forth_fwm_process} (d), showing the spectra of the
components reveals an interesting feature.  Namely one can spot
oscillations of the amplitude of pump wavepacket and we attribute
them to the self phase modulation which was mentioned above as the
trivial solution $Q=0$ (or $q=0$) of the Eq.~(\ref{eq:ss}).
\begin{figure}[th]
\centering
\includegraphics[width=0.90\columnwidth]{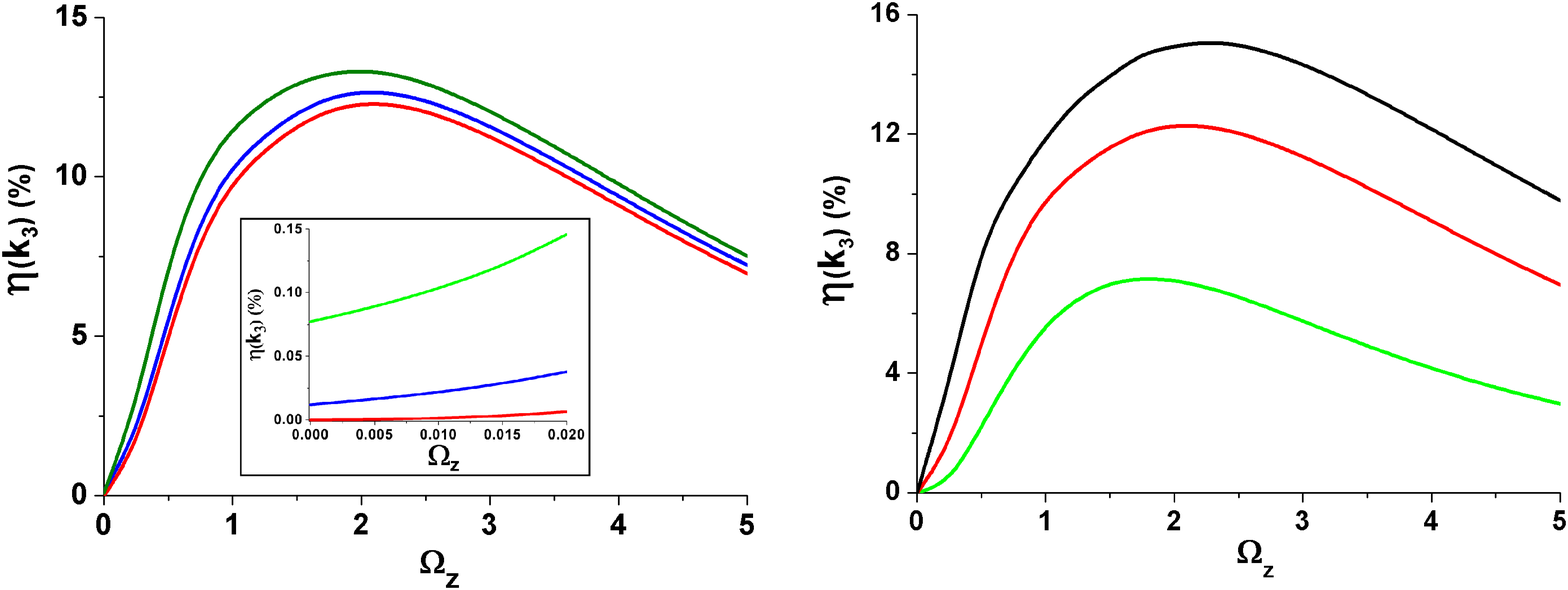}
\caption{Efficiency of the wave generation versus $\Omega_z$ in the
configuration 1 (see Table~\ref{tab:approx} and
Fig.~\ref{fig:first_fwm_process}). In both panels the nonlinearities
($g$, $\overline{g}$ and $\Delta g$) are fixed. In the left panel,
we choose $g=0.8$ and show curves corresponding to $\Delta g/g=0$
(red), $=0.02$ (blue), and $=0.05$ (green). In the right panel, we
fixed $\Delta g/g=0$. The  curves represent three cases: $g=0.5$
(green), $=0.8$ (red) and $=1$ (black). Other parameters are the
same as given in Fig.~\ref{fig:first_fwm_process} for the first
configuration (the other configurations will be similar). The
inset in the left panel magnifies the region of small $\Omega_z$
illustrating the effect of the broken SU(2) symmetry in the FWM
process (see the text).} \label{fig:collective}
\end{figure}

For completeness we present the fourth configuration in the case
when the phase matching allows simultaneously for  \textit{two} FWM
processes. In this last simulation we used parameters corresponding
to the region, where the phase matching equation supports four
different root: $q_1$, $q_2$, $-q_1$ and $-q_2$. Having these roots
in hand, we define momenta of the two sets of probe waves in the
following way:
\begin{eqnarray}
k_2=k_1+q_1, \qquad  k_3=k_1-q_1 \nonumber \\
k_4=k_1+q_2, \qquad k_5=k_1-q_2. \nonumber
\end{eqnarray}
In present case dynamics of the FWM process involves waves with the
amplitudes: $A_1$ as a pump, and two pairs of probs  $A_2$ and
$A_3$,  as well as  $A_4$ and  $A_5$ (by convention a mode with $Aj$
has momentum $k_j$, where $j=1,...,5$).  In numerical simulations we
initiate the dynamics by putting  $A_1=1$, $A_3=A_5=0.2$, and  $A_2=
A_4=0$ [see Eq.(\ref{eq:initial}); the values of the rest of
parameters are listed in the Fig.~\ref{fig:forth_fwm_process4modes}
caption]. In the nonlinear evolution the pump is interacting with
both sets of the probe wavepackets creating two new waves $A_2$ and
$A_4$ with momenta $k_2$ and $k_4$ respectively, in two simultaneous
FWM processes. Figure~\ref{fig:forth_fwm_process4modes} shows two
components of the spinor wavefunctions in the configuration [(a) and
(b)] and momentum [(c) and (d)] spaces, where we can clearly see two
sets of probe waves. This time we can not refer to Fig.~\ref{fig:set
of GVs}, since the values of parameters were slightly different from
those used for panel(c), but for the sake of clarifications we
propose to look at panel (d) of the Fig.~\ref{fig:123_cases}. Due to
the strong spreading there is substantial overlap of wavepackets,
but when we look at them in the momentum space all peaks can be
clearly identified. Close inspection reveals also oscillations on
the pump wave due to the self-phase modulation mentioned above.

In general finding optimal conditions to observe effective FWM
process is not an easy task especially taking into account that the problem is multi-parametric and the values of parameters have to be
carefully selected. In order to support this statement we have chosen the Zeeman splitting $\Omega_z$ as a control parameter and in Fig.~\ref{fig:collective} we present the efficiency (unit of per cent) of
FWM defined as
\begin{equation}
\eta(k_3)=\frac{\widetilde{N}(k_3)}{N}\cdot 100\%,
\label{eq:efficiency}
\end{equation}
where $N$ is total number of atoms in the system and the number of atoms
corresponding to the generated wave $\widetilde{N}(k_3)$ is
evaluated after long enough evolution time, when the wave packets
are well separated. The efficiency is calculated for the FWM conversion in the process identified above as configuration 1 for different for different types of the nonlinearity parameterized by the quantities $\overline{g}=(g_1+g_2)/2$ and $\Delta g=(g_1-g_2)/2$. In all of the
calculations presented in the Fig.~\ref{fig:collective} we took $g=\overline{g}$.

In both panels of the Fig.~\ref{fig:collective} we vary $\Omega_z$
while keeping the rest of the parameters fixed (see
Fig.~\ref{fig:first_fwm_process} for the exact values of all
parameters). In the left panel different curves correspond to
different values of the ratio $\Delta g/g$ with $g=0.8$, while in
the right panel $\Delta g$ is equal to zero and different curves
were obtained choosing different values of the interaction strength
$g$. In both panels we observe non-monotonic behavior of the
efficiency on $\Omega_z$. At $g_1=g_2=g$ (the right panel and the
red curve in the left panel) we find that the efficiency goes to
zero in the limits $\Omega_z\to0$ and $\Omega_z\to \infty$ having
maximum at some finite value of the Zeeman splitting (notice that
both these limits correspond to the integrable cases~\cite{KKMS}). Such
behavior may be explained by the fact that the case of all equal
nonlinearities corresponds to the SU(2) symmetric (also known as
Manakov~\cite{Manakov}) nonlinearity. Thus at $\Omega_z=0$ the
linear Hamiltonian (\ref{hamiltonian}) can be diagonalized by the
global rotation. An approximate diagonalization can be also
performed in the formal limit $\Omega_z\to \infty$ at $\Omega_x$
fixed, which after rescaling can be viewed as the
limit of $\Omega_x\to 0$ at $\Omega_{z}$ fixed. Next we recall, that
for a 1D NLS equation the matching conditions cannot be satisfied,
this meaning that no FWM processes involving eigenstates from the
same branch can be observed. On the other hand, the SU(2)
nonlinearity does not support FWM processes where states from the
different branches are involved. This can be easily seen if we rewrite
the Manakov nonlinearity in the form $G(\bPsi)\equiv \bPsi^\dagger\bPsi$.
Indeed, since the states belonging to different branches of the
spectrum are mutually orthogonal, this nonlinearity does not support
mixing of mentioned states, i.e. the respective frequency conversion
processes is not phase matched. On the other hand, when the nonlinear coefficients are
not exactly equal (blue and green lines in the left panel of
Fig.~\ref{fig:collective}) we observe that even at $\Omega_{z}\to 0$
the efficiency does not vanish, although becomes very small, as
shown in the inset in the left panel of Fig.~\ref{fig:collective}.

\section{Conclusions}
In our study we analyzed the four-wave mixing process in
Bose-Einstein condensates with spin-orbit coupling. We found all
phase matched configurations for degenerate case where two identical
initial states interact with two probe ones. We performed numerical
simulations to illustrate the dynamics in which we seeded one of the
probe and observed stimulated growth of the latter combined with
resonant generation of extra waves. We found unique conditions where
both probe waves have smaller group velocity than pump wave, and
also reported the case when two FWM process can occur
simultaneously.

\section{Acknowledgments}
The work was supported by the Polish National Science Centre
2016/22/M/ST2/00261 (M.T.), Vietnam National Foundation for Science and Technology Development (NAFOSTED) under grant number
103.01-2017.55 (N.V.H.), and Portuguese Foundation for Science and Technology (FCT) under Contract no. UIDB/00618/2020 (V.V.K.).

\end{document}